\documentclass{elsart}
\usepackage{amsfonts,amsmath}
\usepackage{verbatim}
\usepackage{epsfig}
\usepackage{graphicx}

\DeclareMathOperator{\eo}{{\it X}}
\DeclareMathOperator{\eoi}{{\it X}}
\DeclareMathOperator{\eoii}{{\it Y}}
\DeclareMathOperator{\dist}{\delta}
\DeclareMathOperator{\facea}{\gamma}
\DeclareMathOperator{\faceb}{\kappa}
\DeclareMathOperator{\facec}{\eta}
\DeclareMathOperator{\faced}{\rho}
\DeclareMathOperator{\vertexv}{{\it v}}
\DeclareMathOperator{\vertexu}{{\it u}}
\DeclareMathOperator{\vertexw}{{\it w}}
\DeclareMathOperator{\mc}{\mathcal{M}}
\DeclareMathOperator{\path}{\gamma}
\DeclareMathOperator{\cond}{\Phi}

 \DeclareMathOperator{\mct}{\widetilde{\mc}}
\DeclareMathOperator{\pot}{\wp}
\DeclareMathOperator{\faces}{\mathcal{F}}

\newenvironment{itemize*}
  {\begin{itemize}%
    \setlength{\topsep}{0pt}
    \setlength{\itemsep}{0pt}%
    \setlength{\parskip}{0pt}
    \setlength{\parsep}{0pt}
  }%
  {\end{itemize}}

\newenvironment{enumerate*}
  {\begin{enumerate}%
    \setlength{\topsep}{0pt}
    \setlength{\itemsep}{0pt}%
    \setlength{\parskip}{0pt}
    \setlength{\parsep}{0pt}
  }%
  {\end{enumerate}}

\begin{document}

\begin{frontmatter}
\title{Sampling Eulerian orientations of triangular lattice graphs}
\author{P\'aid\'i Creed}
\ead{p.creed@ed.ac.uk}
\address{School of Informatics\\
        University of Edinburgh\\
        Edinburgh EH9 3JZ\\ 
        Scotland}

\begin{abstract}
We consider the problem of sampling from the uniform distribution on
the set of Eulerian orientations of subgraphs of the triangular
lattice. Although it is known that this can be achieved in polynomial time
for any 
graph, the algorithm studied here is more natural in the context of
planar Eulerian graphs. We analyse the mixing time of a Markov chain
on the Eulerian orientations of a planar graph which moves between
orientations by reversing the edges of directed faces. Using
{path coupling} and the {comparison method} we obtain a
polynomial upper bound on the mixing time of this chain for any
{solid subgraph} of the triangular lattice. By considering the
{conductance} of the chain we show that there exist 
{subgraphs with holes} for which the chain will always take an
exponential amount of time to converge. Finally, as an additional
justification for studying a Markov chain on the set of Eulerian
orientations of planar graphs, we show that the problem of counting
Eulerian orientations remains \#P-complete when restricted to planar
graphs.    

A preliminary version of this work appeared as an extended abstract in
the 2nd Algorithms and Complexity in Durham workshop.

\end{abstract}

\end{frontmatter}

\section{Introduction}\label{sec:intro}
Let $G=(V,E)$ be an Eulerian graph, that is, a graph with all vertices
of even degree. An Eulerian orientation of $G$ is an orientation of
the edges of $G$ such that for every vertex $\vertexv$ the number of edges
oriented towards $\vertexv$ is equal to the number oriented away from
$\vertexv$: $deg_{in}(\vertexv)=deg_{out}(\vertexv)$. It is well-known
that the problem of finding an Eulerian orientation of an Eulerian
graph can be solved efficiently; in this paper we consider
the problem of sampling the set of Eulerian orientations of a
planar Eulerian graph, that is, the problem of generating an Eulerian
orientation from a distribution that is close to uniform. We focus on
the \emph{Markov chain Monte Carlo} method, a standard approach to random
sampling of combinatorial structures. The Markov chain we study is the
most natural chain whose state space is the set of Eulerian
orientations of a planar graph. To move from one Eulerian
orientation to another the chain randomly selects a face of the
graph. If the edges of this face form a directed cycle in the original
Eulerian orientation the chain reverses the orientation of its
edges. We will hereafter refer to this chain as the
\emph{face-reversal} chain.    

Markov chain simulation is generally only useful when we know the
chain is \emph{rapidly mixing}, that is, when the number of steps
required to get within variation distance $\epsilon$ of the stationary
distribution is bounded from above by a polynomial in the size of $G$
and $\epsilon^{-1}$. 

The problems of sampling and counting are closely related; indeed,
almost all approximate counting algorithms rely on the existence of an
efficient sampling algorithm. The \#P-completeness of counting
Eulerian orientations of a graph was established by Mihail \& Winkler
\cite{MihailW:ALGO96}, thus motivating the study of rapidly mixing
Markov chains for this problem.  More recently, Felsner \& Zickfeld
\cite{FelsnerZ:NoAO} have obtained upper and lower bounds on the number
of Eulerian orientations of any planar map in the more general context
of $\alpha$-orientations.    

Mihail \& Winkler \cite{MihailW:ALGO96} demonstrated that the problem
of sampling from the Eulerian orientations of any graph can be reduced to
the problem of sampling a perfect matching of a specially constructed
bipartite graph. Sampling a perfect matching of this bipartite graph
can be achieved in polynomial time using a Markov chain shown to be
rapidly mixing by Jerrum \& Sinclair \cite{JerrumS:SIAMCOMP89} (see
also \cite{JerrumSV:JACM01,BezakovaSVV:SODA06}). Nevertheless,
studying the face-reversal chain is still of interest as it is the
more natural approach in the context of planar graphs. In particular,
in this paper we show that the face-reversal chain provides a more
efficient sampling algorithm for the case of solid subgraphs of the
triangular lattice, which correspond to configurations of the
20-vertex ice model studied by statistical physicists
\cite{Baxter:JMP10}. 

The problem of counting Eulerian orientations of
several planar lattices has been studied in statistical physics as it
corresponds to evaluating the partition function $Z_{ICE}$ of certain
\emph{ice models}
\cite{Baxter:SolvedModels,Eloranta:ArchimedeanIce}. In particular,   
Baxter \cite{Baxter:JMP10} has found an asymptotic estimate for the
number of Eulerian orientations of a grid-like section of the
triangular lattice.  

Goldberg et al. \cite{GoldbergMP:RSA04} and Luby et al.
\cite{LubyRS:SIAMCOMP01} showed that the face-reversal chain is
rapidly mixing on the set of Eulerian orientations of rectangular
sections of the square lattice in the cases of fixed and free boundary
conditions, respectively. Both these proofs followed a similar
pattern. First they extended the chain with extra transitions. Then a
coupling argument was used to find a bound on the mixing time of this
extended chain. Finally, the comparison technique of Diaconis \& Saloff-Coste
\cite{DiaconisS-C:AnnalsAppProb:93} was applied to infer rapid mixing
of the face-reversal chain. Fehrenbach \& R\"uschendorf
\cite{FehrenbachR:ATCP04} studied the mixing time of the face-reversal
chain on the set of Eulerian orientations of the triangular
lattice. Following the approach of
\cite{LubyRS:SIAMCOMP01,GoldbergMP:RSA04} they defined an extension of
the face-reversal chain, and attempted to use the path coupling
technique of \cite{BubleyD:FOCS97} to show that it is rapidly
mixing. However, their analysis does not seem to be correct so this
problem is still open. 

In this paper we follow a similar approach to
\cite{LubyRS:SIAMCOMP01,GoldbergMP:RSA04} 
to show that the face-reversal chain is rapidly mixing on the set of
Eulerian orientations of \emph{solid} subgraphs of the triangular
lattice. In addition to this positive result, we exhibit an infinite
family of subgraphs of the triangular lattice for which this chain is
not rapidly mixing. Finally, we include a proof that the problem of
counting Eulerian orientations remains \#P-complete when restricted 
to planar graphs as an additional justification for studying the
mixing time of the face-reversal Markov chain. 
 
The remainder of the paper is laid out as follows:\\
In \S\ref{sec:bg} we define the notation we use for Markov chains and
Eulerian orientations, along with an alternative way to view Eulerian
orientations (due to Felsner \cite{Felsner:EJCOMB04}) which proves
useful for our analysis. We also describe the machinery that we will
use to prove results about the mixing time of the chains we
study. \S\ref{sec:mc} contains a description of two Markov chains, the
face-reversal Markov chain that is the chief interest of this paper,
and an extension of this chain which we call the the
\emph{tower-moves} chain. \S\ref{sec:mixing} contains the
details of the path-coupling argument we use to show that this 
extended chain mixes rapidly for solid
subgraphs of the triangular lattice. This is then used to infer rapid
mixing of the face-reversal chain via the comparison method of
Diaconis \& Saloff-Coste \cite{DiaconisS-C:AnnalsAppProb:93} in
\S\ref{sec:comp}. In \S\ref{sec:torpid} we show that there exist
subgraphs of the triangular lattice \emph{with holes} for which the
face-reversal chain will always take an exponential amount of time to
converge. Finally, \S\ref{sec:planarhard} contains the proof that
exact counting of Eulerian orientations remains hard when restricted
to planar graphs.  

\section{Background}\label{sec:bg}
In this section we summarise the techniques and theory used in our
analysis. 

A discrete-time Markov chain with transition probability matrix $P$
defined on a finite state space $\Omega$ is called \emph{ergodic} if
it is both {\em aperiodic} and {\em irreducible}. Any ergodic Markov
chain has a unique stationary distribution $\pi$. Moreover, if $P$ is
symmetric then $\pi$ is uniform over $\Omega$.  Given two
probability distributions, $\rho$ and $\mu$, on $\Omega$ the
\emph{variation distance} is defined as  
$$|| \rho-\mu ||_{TV} = \displaystyle\mathop{\sup}_{A \subset \Omega}
| \rho(A)-\mu(A) |\,.$$ 
The mixing time is a measure of the number of steps taken by a Markov
chain to get close to its stationary distribution. This is defined as 
$$\tau(\epsilon) = \displaystyle\mathop{\sup}_{x \in \Omega} \min
\{t : ||P^t(x,\cdot) - \pi ||_{TV} \leq \epsilon\}\,.$$
A Markov chain is said to be rapidly mixing if $\tau(\epsilon)$ is
bounded above by some polynomial in the size of the elements of
$\Omega$ and in $\epsilon^{-1}$. For example, in this paper we use the
number of faces to measure the size of the problem.

Coupling is a standard technique for proving upper bounds on the
mixing time of Markov chains
\cite{Aldous:PROBXVII:83,BubleyD:FOCS97}. A coupling of a chain
$\mathcal{M}$ is a stochastic process $(X_t,Y_t)_{t \geq 0}$ on
$\Omega \times \Omega$ such that each of the marginal distributions of  
$(X_t)_{t \geq 0}$ and $(Y_t)_{t \geq 0}$ are a faithful copy of the
original Markov chain. To bound the mixing time via coupling we use
the coupling inequality \cite{Aldous:PROBXVII:83}, which states that
the variation distance between $\pi$ and the distribution at time $t$
is bounded above by the probability of any coupling coalescing by time
$t$, ie.   
$$|| P^t - \pi||_{TV} \leq
\displaystyle\mathop{\sup}_{X_0,Y_0}Pr[X_t \neq Y_t]\,.$$
Therefore, in order to obtain a polynomial bound on the mixing time it
suffices to construct a coupling which will have coalesced (with high
probability) after a polynomial number of steps. This can be proven by
showing that the coupling causes all pairs of states to ``move
together'' under some measure of distance.  

We will use a simplified variation of coupling, due to
\cite{BubleyD:FOCS97}, known as \emph{path coupling}. This involves
defining a coupling $(X_t,Y_t)$ by considering a \emph{path}
$X_t=Z_0,Z_1,\ldots,Z_r=Y_t$ between $X_t$ and $Y_t$ where each pair
$(Z_i,Z_{i+1})$ is adjacent in the Markov chain and the path is a
shortest path between $X$ and $Y$. This allows us to
restrict our attention to the pairs of states which are adjacent in
the chain, as shown by the following theorem:  
\begin{thm}[Bubley \& Dyer \cite{BubleyD:FOCS97}]\label{thm:path}
Let $\mc$ be an ergodic Markov chain with state space $\Omega$ and let $\delta$
be an integer valued metric defined on $\Omega\times\Omega$ which
takes values in $\{0,\ldots,D\}$. Let $S$ be a subset of
$\Omega\times\Omega$ such that for all $(X,Y) \in 
\Omega\times\Omega$ there exists a path
$$X = Z_0,Z_1,\ldots,Z_r=Y$$
between $X$ and $Y$ such that $(Z_i,Z_{i+1}) \in S$ for $0 \leq i < r$
and 
$$\sum_{i=0}^{r-1}\delta(Z_i,Z_{i+1}) = \delta(X,Y)\,.$$ 
Now suppose $(X_t,Y_t)$ is a coupling of $\mc$ defined on $S$. If
there exists $\beta \leq 1$ such that for all $(X,Y) \in S $ 
$$\mathbb{E}[\delta(X_{t+1},Y_{t+1}) | (X_t,Y_t) = (X,Y)] \leq
\beta\delta(X_t,Y_t)$$  
then this coupling can be extended to a coupling $(X_t,Y_t)$ defined
on the whole of $\Omega\times\Omega$ such that  
$$\mathbb{E}[\delta(X_{t+1},Y_{t+1})] \leq \beta\delta(X_t,Y_t)\,.$$ 
Moreover, if $\beta < 1$ then $\tau(\epsilon) \leq
\frac{\log(D\epsilon^{-1})}{1-\beta}$. 
\end{thm}

If $\beta=1$ in Theorem~\ref{thm:path} then in order to use standard
path coupling techniques it must be shown that the variance of the
distance between \emph{any} two states after one step of the coupling
can be bounded away from $0$ \cite{BubleyD:FOCS97}. However, a recent
result has removed this condition: 

\begin{thm}[Bordewich \& Dyer \cite{BordewichD:Equality}]\label{thm:pc}
Suppose we have a path coupling $(X_t,Y_t)$ for an ergodic Markov
chain $\mc$ with distance metric
$\dist:\Omega\times\Omega\rightarrow\mathbb{N}$, and suppose $S
\subset \Omega\times\Omega$ is the set of pairs of states at distance 1. Let
$t(\epsilon) = \lceil p^{-1}eD^2\rceil\lceil\log\epsilon^{-1}\rceil$,
where p is the minimum transition probability between pairs of states
in S, D is the maximum distance between any pair of states, and let 
$t^\star=Bin(t(\epsilon),(1+p)^{-1})$, where $Bin$ denotes a
binomially distributed random variable. If $\beta\leq 1$ for the
coupling, then at the random time $t^\star$ the Markov chain is
within $\epsilon$ of the stationary distribution, in total variation
distance.  
\end{thm} 
In \S\ref{sec:mixing} we will apply Theorem~\ref{thm:pc} to show that an
extension of the face-reversal chain is rapidly mixing on any solid
subgraph of the triangular lattice.

To prove a lower bound on the mixing time of a Markov chain we rely on
the notion of conductance. For any non-empty set $S \subset \Omega$
the conductance $\cond(S)$ of $S$ is defined to be
$$\cond(S) = \frac{\sum_{x \in S, y \in \Omega\backslash S}\pi(x)P(x,y)}
                  {\pi(S)}\,.$$

It is well-known that conductance is inversely related to the mixing
time (see eg. \cite{SinclairJ:InfComp89,LuczakV:JDA05});
$$\tau(e^{-1}) \geq \frac{1}{2\min_{0 < \pi(S) < 1/2}\cond(S)}\,.$$
Thus, if $n$ is some measure of the size of the elements of $\Omega$
then by finding an upper bound on the conductance of a Markov chain
that is exponentially small in $n$ we can conclude that the chain is
not rapidly mixing. Conventionally, a chain is said to be
\emph{torpidly mixing} when it is known that it is not rapidly mixing.  

Defining the boundary of a set $S$ as
$$\partial S = \{x \in S: P(x,y) > 0 \textnormal{ for some } y \in
\Omega\backslash S\}$$
we get
$$\cond(S) \leq \pi(\partial S)/\pi(S)\,.$$ 
Hence, to show that a chain is torpidly mixing it suffices to find a set
for which this last expression is bounded above by some exponentially
small function. This is encapsulated in the following theorem, taken
from \cite{LuczakV:JDA05}: 
\begin{thm}\label{thm:torpid}
If, for some $S \subset \Omega$ satisfying $0 < |S| \leq |\Omega|/2$, the
ratio $\pi(\partial S)/ \pi(S)$ is exponentially small in the size of
the elements of $\Omega$, then the Markov chain is torpidly mixing. 
\end{thm} 
We will use the above theorem in \S\ref{sec:torpid} to show that there
exists subgraphs of the triangular lattice for which the face-reversal chain
is torpidly mixing. 

Let $G=(V,E)$ be a planar graph and $\faces(G)$ be the set of bounded
faces in some planar embedding of $G$. We will use $f$ to
denote the number of elements in $\faces(G)$ and $EO(G)$ to denote the
set of Eulerian orientations of $G$. For any simple cycle $C \subset
E$, $Int(C)$ is defined to be the set of faces on the interior of
$C$. A face is said to be  directed in an orientation of $G$ if its
boundary edges form a directed cycle. Felsner \cite{Felsner:EJCOMB04}
showed that it is possible to convert any Eulerian orientation of $G$
into another by performing a sequence of reversals of the edges of
directed faces. Furthermore, a partial order was defined 
on $EO(G)$ by letting $\eoi \prec \eoii$ if $\eoi$ can be obtained
from $\eoii$ by performing a sequence of reversals of clockwise
directed faces. This order has a unique  maximum
(resp. minimum) element: the unique Eulerian orientation with 
no clockwise (resp. counter-clockwise) cycles. Felsner proved this
order forms a finite distributive lattice by giving a bijection between
the Eulerian orientations of a planar graph and a set of functions
of the form $EO(G)\rightarrow\mathbb{N}$ called $\alpha$-potentials. To 
define these  functions Felsner used a partial order on $\faces(G)$:
$\faceb \prec_{\faces} \facec$ if $\faceb$ and $\facec$ share an edge
and that edge is clockwise on $\faceb$ in the minimum orientation. The
$\alpha$-potentials are then defined as the set of functions
$\pot_{\eo}:\faces(G)\rightarrow \mathbb{N}$ such that:  
\begin{align}
\textnormal{$\faceb$ and $\facec$ share an edge} &\Rightarrow
|\pot_{\eo}(\faceb)-\pot_{\eo}(\facec)| \leq 1 \label{expr:pot1}\\ 
\textnormal{$\faceb$ is on the boundary} &\Rightarrow \pot_{\eo}(\faceb) \leq 1
\label{expr:pot2} \\ 
\faceb \prec_{\faces} \facec &\Rightarrow \pot_{\eo}(\faceb) \leq
\pot_{\eo}(\facec) \label{expr:pot3}  
\end{align}
 
The bijection is given as follows: let $\pot_{\eo}(\faceb)$ equal the
number of times $\faceb$ is reversed on any shortest path from $\eo_{\min}$
to $\eo$. 

In this paper we use Felsner's results to show that the face-reversal
Markov chain  is irreducible and to
define a natural distance metric on the set of Eulerian orientations
of a planar graph. For the rest of this paper we will refer to this
lattice as the \emph{Felsner lattice} of a planar graph G, and denote
it by $Fels(G)$.    

\begin{figure}[h]
\begin{center}
\includegraphics[width=6cm,height=5cm]{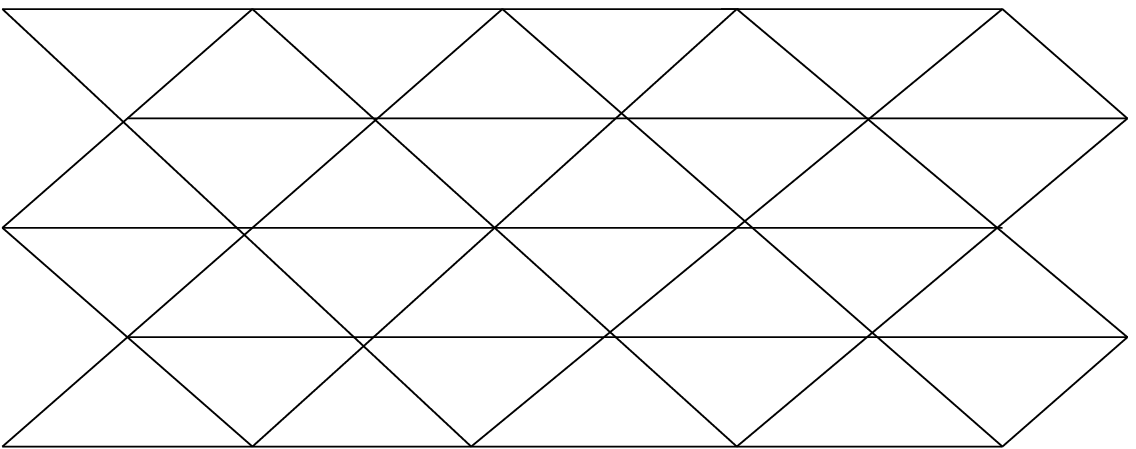}
\end{center}
\caption{A section of the triangular lattice}
\label{fig:triangle}
\end{figure}

The (infinite) triangular lattice is a structure studied by members of
the statistical physics community, 
eg.
\cite{Baxter:JMP10,Baxter:SolvedModels,Eloranta:ArchimedeanIce}. This
is the infinite graph with the following description:
\begin{align*}
V&=\{v_{i,j} : i,j \in \mathbb{Z}\}\\
E&=\{\{v_{i,j},v_{i+1,j}\},\{v_{i,j},v_{i,j+1}\},\{v_{i,j},v_{i+1,j-1}\}
: i,j \in \mathbb{Z}\}\,.
\end{align*}
This graph can be embedded in the plane as shown in
Figure~\ref{fig:triangle}. In this paper we focus on two
\emph{finite} subgraphs of this lattice:    
\begin{enumerate}
\item A \emph{solid subgraph} is a graph which can be defined by
  specifying a cycle in the triangular lattice as the boundary and
  taking everything on its interior.
\item A \emph{subgraph with holes} is defined similarly to a
  \emph{solid subgraph} except we do not take everything on the
  interior of the cycle.
\end{enumerate}
\S\ref{sec:mixing} and \S\ref{sec:comp} are devoted to showing that
the face-reversal chain mixes rapidly on the set of Eulerian
orientations of any Eulerian solid subgraph. In \S\ref{sec:torpid} we
show that this is not true in the case of subgraphs with holes. We
prove this negative result by exhibiting an infinite family of Eulerian
subgraphs with holes on which the face-reversal chain is torpidly
mixing.  

\section{The Markov chains}\label{sec:mc}

We now define the face-reversal Markov chain on the set of Eulerian
orientations of any Eulerian planar graph. We use $X$ and $X^\prime$ to
denote the states of the chain before and after each step.    
  
{\bf One step of the chain $\mc$}
\begin{enumerate*}
\item With probability $1/2$, set $X'=X$
\item With the remaining probability
      \begin{enumerate*}
      \item Choose $\facea \in \faces(G)$ u.a.r.
      \item If $\facea$ is directed then obtain $X'$ from $X$ by reversing
      the orientation of all the edges in $\facea$.      
      \item Otherwise, set $X'=X$
      \end{enumerate*}
\end{enumerate*}

\emph{Irreducibility} of this chain follows from the fact that the
underlying graph of the chain is the cover graph of $Fels(G)$. The
holding probability of $1/2$ guarantees \emph{aperiodicity}, whence
the chain is ergodic and converges to a unique stationary distribution
$\pi$. Moreover, the chain is symmetric so $\pi$ is the uniform
distribution on the set of Eulerian orientations of $G$.

In order to apply the path coupling theorem we will need extend this
chain with extra moves in the style of
\cite{LubyRS:SIAMCOMP01,GoldbergMP:RSA04}. These
moves allow us to couple with $\beta \leq 1$, something which does not
seem to be possible for the original chain with the metric we use
(to be defined in \S\ref{sec:mixing}). 

\begin{defn}\label{defn:tower}
Let $G$ be an Eulerian planar graph, $\eo$ an Eulerian orientation of $G$,
and $\facea$ a face of $G$. We say $\facea$ is \emph{almost-directed}
in $\eo$ if all but one of the edges of $\facea$ have a common
direction. We call the edge with the disagreeing direction the
\emph{blocking edge} of $\facea$. For faces $\faceb,\facec \in
\faces(G)$ we say there is a tower $T$ starting at $\faceb$ and ending
at $\facec$ in an orientation $\eo$ if there is a set of adjacent
faces $\faceb=\facea_1,\ldots,\facea_h=\facec$ such that  
\begin{itemize*}
\item $\facea_i$ is almost directed in $\eo$, and the blocking edge of
      $\facea_i$ is the one shared with $\facea_{i+1}$ for $1 \leq i
      \leq h-1$.        
\item $\facea_h$ is a directed face in $\eo$
\end{itemize*}
We say $h$ is the \emph{length} of the tower.
\end{defn}
Observe that the definition of a tower implies that
$C=\bigoplus_{1\leq i\leq h}E(\facea_i)$ is a directed cycle in
$\eo$. We say that a tower is clockwise (resp. counter-clockwise) in
an orientation if this cycle is clockwise (resp. counter-clockwise) in
the orientation. We call $\facea_h$ and $\facea_1$ 
the top and bottom of the tower, and refer to the right
and left sides of the towers in terms of a walk from the bottom to the
top. It follows that in a clockwise tower the internal edges are all
directed from the right to the left, and vice-versa for
counter-clockwise towers. 

Let $\eo \in EO(G)$ and let $\facea \in \faces(G)$ be almost directed
in $\eo$. If there is a tower in $\eo$ starting at $\facea$ we can
find it by walking along the faces of $G$, starting at $\facea$ and
choosing the face sharing the blocking edge with the current face at
each step. If at any point we reach a directed face then we have
found a tower. If we reach an undirected face which is not almost
directed, or whose disagreeing edge lies on the boundary of the graph,
or whose blocking edge is the same edge as the blocking edge of the
previous face, then there cannot exist a tower starting at
$\facea$. To see that this process terminates (ie. does not wrap
around on itself) consider the case where we reach the starting face
again. For this to occur we must have two cycles $C_1$ and $C_2$,
where  
$$Int(C_2) \cup \{\facea_i | 1 \leq i \leq r\} = Int(C_1)\quad
and\quad \facea_i \notin Int(C_2)\,,$$ 
with all the edges connecting $C_1$ to $C_2$ oriented towards $C_1$ or
$C_2$, something which is impossible in an Eulerian orientation.
In particular, all towers in an Eulerian
orientation of a solid subgraph of the triangular lattice are linear
since any vertex at which a tower bends will be unbalanced.  

We now define the extended \emph{tower-moves} chain. The definition
includes an undetermined probability $p_T$ which will be fixed later.

{\bf One step of the Markov chain $\mct$}
\begin{enumerate*}
\item With probability $1/2$, set $X'=X$
\item With the remaining probability
      \begin{enumerate*}
      \item Choose $\facea \in \faces(G)$ u.a.r.
      \item If $\facea$ is directed then obtain $X'$ from $X$ by reversing
      the orientation of all the edges in $\facea$.
      \item Otherwise, if there is a length $h$ tower $T=(\facea_i)_{1
      \leq i \leq h}$ with $\facea_1=\facea$ then let $C=\bigoplus_{1
      \leq i \leq h}F_i$. With probability $p_T$ obtain $X'$ from $X$
      by reversing all the edges of $C$. 
      \item Otherwise, set $X'=X$
      \end{enumerate*}
\end{enumerate*}
This type of chain has been used to extend the face reversal
chain in the past, see
\cite{LubyRS:SIAMCOMP01,GoldbergMP:RSA04}. The ergodicity of this 
chain is inherited from the ergodicity of $\mc$, since every
transition in $\mc$ is also a transition in $\mct$. As long as $p_T$
is chosen so that the probability of reversing a tower is independent
of whether it is a clockwise or a counter-clockwise tower, $\mct$
converges to the uniform distribution. To see this suppose $\eoi$ can
be obtained from $\eoii$ by reversing a clockwise tower $T$. But then
we can obtain $\eoii$ from $\eoi$ by reversing a counter-clockwise
tower containing the same faces as $T$, so
$P(\eoi,\eoii)=P(\eoii,\eoi)$. Hence, the stationary distribution of
$\mct$ is also the uniform distribution on the set of Eulerian
orientations of $G$.     

\section{Rapid Mixing of $\mct$ on solid subgraphs}\label{sec:mixing}
In this section we use the path coupling technique of Bubley \& Dyer
\cite{BubleyD:FOCS97} to show that the Markov chain $\mct$ is rapidly
mixing on $EO(G)$ when $G$ is a solid subgraph of the triangular
lattice and $p_T$ is chosen appropriately. This is not the first
result regarding the mixing time of this type of chain. $\mct$ has
been shown to be rapidly mixing on the square lattice (using different
$p_T$ values) with different types of boundary conditions in 
\cite{LubyRS:SIAMCOMP01,GoldbergMP:RSA04}.   
Fehrenbach \& R\"uschendorf \cite{FehrenbachR:ATCP04} attempted to
give a proof of rapid mixing for a related chain (in which only towers
of length $2$ were used). However, there are errors in the analysis
of \cite{FehrenbachR:ATCP04}, and it seems as though their path coupling 
analysis cannot be fixed for the particular chain used. 

Note that rapid mixing proofs for this chain are dependent on the
correct choice of the probabilities $p_T$. For example, the proof of
\cite{GoldbergMP:RSA04} sets $p_T$ to $1/4h$ if the tower runs along
the boundary, and $1/2h$ otherwise, where $h$ is the length
of the tower $T$. We now state the main result of this section but
defer the proof until we have presented some useful lemmas. 
\begin{thm}\label{thm:mixing}
For any solid subgraph of the triangular lattice the Markov chain
$\mct$ is rapidly mixing with 
$$\tau_{\mct}(\epsilon) \in O(f^4\log\epsilon^{-1}),$$
when $p_T=1/3h$ for all towers $T$ of length $h$.  
\end{thm}

In our application of path coupling to proving rapid mixing of $\mct$
we will need a metric on the set of Eulerian orientations of a planar 
graph $G$. We define $\phi(G)$ to be the covering graph of
$Fels(G)$. With this definition, $\phi(G)$ is a connected graph with
edge set
$$\{\{\eoi,\eoii\} : \textnormal{$\eoi$ can be obtained from $\eoii$ by 
  reversing a single face}\}\,.$$
We now define our metric $\dist:\Omega\times\Omega
\rightarrow \{0,1,\ldots,D\}$:  
$$\dist(\eoi,\eoii) = \min \{|\path_{\eoi,\eoii}|:
  \path_{\eoi,\eoii} \textnormal{is a path from $\eoi$ to $\eoii$ in
  } \phi(G)\}\,.$$  
In particular $\dist(\eoi,\eoii)=1$ if $\eoi$ and $\eoii$ differ on
a single face.  
Let $\eo_{\min}$ and $\eo_{\max}$ denote the unique minimum and maximum
elements of $Fels(G)$. The fact that $Fels(G)$
is a distributive lattice implies $D=\dist(\eo_{\min},\eo_{\max})$,
since we can always find a path of length $D$ between two states
$\eoi$ and $\eoii$ either by going down from $\eoi$ to $\eo_{min}$ and then
up to $\eoii$, or by taking a similar path through $\eo_{\max}$.
 
\begin{figure}
\begin{center}
\includegraphics{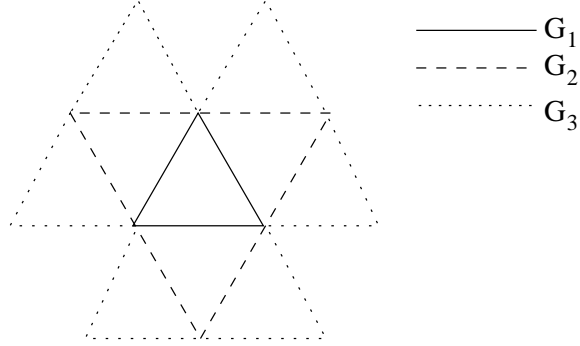}
\end{center}
\label{fig:gk}
\caption{$G_1$, $G_2$, and $G_3$ laid over each other}
\end{figure}
\begin{lem}\label{lem:dist}
The maximum distance between any pair of Eulerian orientations of a
solid subgraph of the triangular lattice $G$ with $f$ bounded faces is
$O(f^\frac{3}{2})$. 
\begin{pf}
From the definition of the bijection between Eulerian orientations and
$\alpha$-potentials given in \S\ref{sec:bg} we can conclude that the
distance between the maximum and minimum Eulerian orientations
is $\sum_{\facea \in \faces(G)}\pot_{\max}(\facea)$. 

(\ref{expr:pot1}) and (\ref{expr:pot2}) imply that
$\pot_{\max}(\facea)$ is exactly the minimum number of edges a line has
to cross to going from $\facea$ to the boundary. Let $G_k$ be the
smallest graph which contains a face $\facea$ with
$\pot_{\max}(\facea)=k$. We can construct $G_k$ inductively, starting
with $G_1=K_3$. To extend $G_k$ to $G_{k+1}$ we add a cycle
containing all the vertices on the boundary of $G_k$ along with any
vertices necessary to ensure that $G_k$ is a solid subgraph of the
triangular lattice, see Figure~\ref{fig:gk} for an example of how this
works. A simple inductive argument shows that the number of faces 
added at each step is $3k$. This implies $|\faces(G_k)| \in
\Theta(k^2)$, so $\pot_{\max}(\facea) \in O(\sqrt{f})$ for any face
$\facea$. 
\end{pf}
\end{lem}

In the following let $\eoi,\eoii$ denote two Eulerian orientations of
$G$ which are adjacent in $\phi(G)$, let $\facea$ be the face on which
they disagree, and let $N(\facea)$ denote the set of faces which share an
edge with $\facea$.  Let $\faceb$ be the face chosen at some step in
the coupling. We then define the coupling as follows:
\begin{itemize*}
\item with probability $1/2-1/2f$, both chains remain unchanged,
\item with probability $1/2f$ chain 1 moves using $\faceb=\facea$, and chain
  2 remains unchanged,
\item with probability $1/2f$ chain 2 moves using $\faceb=\facea$, and chain
  1 remains unchanged,
\item for each $\faceb \neq \facea$, with probability $1/2f$ both chains
  attempt to move using $\faceb$. 
\end{itemize*}
In order to apply the path coupling theorem we need
to show that the expected distance between the two chains does not
increase in a single step of the coupling. To do this we need to consider
which choice of faces will cause the distance to increase, which will
leave the distance unchanged, and which will cause the distance to
decrease. We say the move at $\faceb$ involves $\facec \in N(\facea)$
if $\faceb=\facec$ or there is a tower starting at $\faceb$ that
contains $\facec$. The distance between the coupled chains may
increase iff $\faceb \neq \facea$ but the move at $\faceb$ involves
some $\facec \in N(\facea)$.  
In the following, we use $\dist_{\facec}$ to denote the maximum total
expected change to the distance between the states of the coupled
chains after a single step resulting from moves involving the face
$\facec$ in either chain. 

\begin{lem}\label{lem:directed}
Suppose $\facec \in N(\facea)$ is a directed face in $\eoi$. Then
$\mathbb{E}[\dist_{\facec}] = \frac{1}{3f}$. 
\begin{pf} Let $\faceb \in \faces(G)$ such that selecting $\faceb$
gives a move which involves $\facec$ in at least one of the coupled
chains. We have two cases to consider.  

{\bf Case $\faceb=\facec$: } Since $\facec$ is a neighbour of
$\facea$, and $\facec$ is directed in $\eoi$, it follows that the
blocking edge of $\facec$ in $\eoii$ is the edge shared with
$\facea$. Then $T=\{\facec,\facea\}$ is a tower of length $2$ in
$\eoii$ with a reversal probability of $\frac{1}{6}$. Hence, the
coupling amounts to reversing $\facec$ in $\eoi$ and $T$ in $\eoii$
with probability $\frac{1}{6}$, and reversing $\facec$ in $\eoi$ but
leaving $\eoii$ unchanged with the remaining probability. The former
results in coalescence, whereas the latter yields a pair of
orientations which are distance $2$ apart.  
Thus, since $\facec$ is reversed in $\eoi$ with probability $1$, the
expected value of $\dist_{\facec}$ is 
$$\frac{1}{2f}\left[(+1)(1-\frac{1}{6}) - 1(\frac{1}{6})\right] =
\frac{1}{3f}\,. $$

{\bf Case $\faceb \neq \facec$:} If the move at $\faceb$ involves
$\facec$ then there is a tower $T$ starting at $\faceb$ in at least
one of $\eoi$ and $\eoii$ which contains $\facec$. Since $\facea$ and 
$\facec$ are both directed in $\eoi$ it follows that there must be a
tower $T_1$ starting at $\faceb$ and ending at $\facec$ in $\eoii$
that does not contain $\facea$, and a tower $T_2=T_1 \cup \{\facea\}$
starting at $\faceb$ in $\eoi$.   
Let $h$ be the length of $T_1$, so $T_1$ is reversed in $\eoi$ with
probability $\frac{1}{3h}$ and $T_2$ is reversed in $\eoii$ with probability
$\frac{1}{3(h+1)}$. Observe that if we  
reverse $T_1$ in $\eoi$ we obtain an orientation which is distance
$h+1$ from $\eoii$, but if we also reverse $T_2$ in $\eoii$ then
we have the same orientation in both chains. Therefore, the expected value of
$\dist_{\facec}$ is  
$$\frac{1}{2f}\left[h\left(\frac{1}{3h}-\frac{1}{3(h+1)}\right) - 
  \frac{1}{3(h+1)}\right] = 0\,.$$ 

Thus, the only face whose selection will result in a move which
involves $\facec$ and affects the distance between the coupled chains
is $\facec$ itself, so $\mathbb{E}[\dist_{\facec}] = 1/3f$.  
\end{pf} 
\end{lem}

\begin{lem}\label{lem:undirected}
If $\facec \in N(\facea)$ is not directed in $\eoi$ or $\eoii$
then $\mathbb{E}[\dist_{\facec}]$ is no more than $\frac{1}{3f}$.
\begin{pf}
Observe that, since $\facec$ is not directed in $\eoi$ or $\eoii$,
the blocking edge of $\facec$ will not be shared with $\facea$ in
either copy and that the blocking edge is different in both. Hence, no
tower can include both $\facec$ and $\facea$. Assuming there is a tower
containing $\facec$ in at least one of the two orientations we have
two cases to consider. If one of the two orientations
has a tower starting at some $\faceb \notin \facea \cup N(\facea)$
that contains $\facec$ then there will be no tower containing
$\facec$ in the other orientation. This is because the tower construction
algorithm described in \S\ref{sec:mc} will reach a pair of faces who
share the same blocking edge ($\facec$ and $\faceb$). Thus, when we
are in this situation we can assume that one of the orientations will
be unchanged after one step of the coupling. In the second case there may
be a tower starting at $\facec$ in either orientation.    

\begin{figure}[h]
\begin{center}
\includegraphics[height=7.5cm,width=8.5cm]{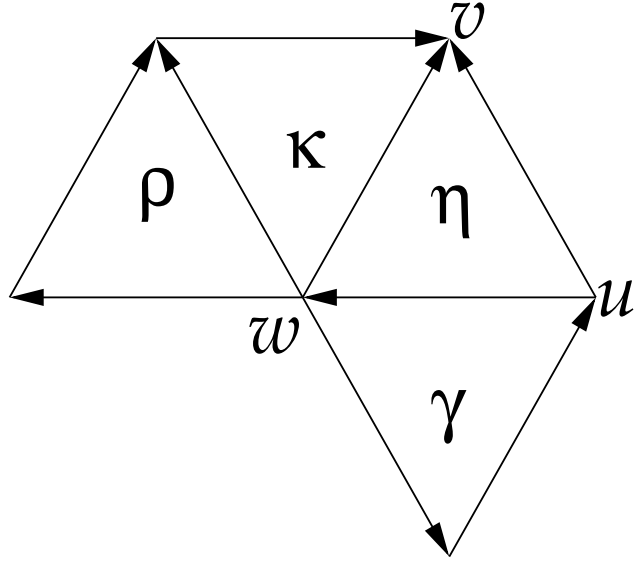}
\end{center}
\label{fig:lemundirected}
\caption{Example from Lemma~\ref{lem:undirected}}
\end{figure}
\begin{list}{{\bf Case \arabic{enumi}:}}{\usecounter{enumi}}
\item Suppose (w.l.o.g.) $\eoi$ is the orientation with a tower containing
$\facec$. Since no move involving $\facec$ is possible in the other
orientation we only need to bound the total expected change to the
distance between $\eoi$ and all $\eoi^{\prime}$ which can be obtained by
making a move involving $\facec$ in $\eoi$.  

We begin by showing that any tower containing $\facec$ in $\eoi$ must
start at $\facec$ or a neighbour of $\facec$. To see this suppose we
have a tower containing $\faced$, $\faceb$, and $\facec$ where
$\faceb$ is a neighbour of $\facec$ and $\faced \in N(\faceb)
\backslash \{\facec\}$. Let $\vertexu,\vertexv,\vertexw$ be the
vertices of $\facec$, and suppose that the edges of $\facec$ are
oriented $\vertexu\vertexv$, $\vertexu\vertexw$, and
$\vertexw\vertexv$. Assume also that $\facec$ shares
$\{\vertexv,\vertexw\}$ with $\faceb$, and $\facec$ shares
$\{\vertexu,\vertexw\}$ with $\facea$. Then $\faced$ must contain
$\vertexw$ (recall all towers are linear). To satisfy the definition
of a tower $\faced$ must have two edges oriented away from
$\vertexw$. But this implies that there are $4$ edges oriented away
from $\vertexw$ in $\eoi$, a contradiction 
(eg. see Figure~\ref{fig:lemundirected}). A similar argument holds for
any other configuration of the edges of $\facec$. Hence, any tower
containing $\facec$ must start at $\faceb$ or $\facec$.    

Let $h$ be the length of the tower starting at $\facec$. Then, in the
coupling, the move in which $\facec$ is the chosen face is made with
probability $\frac{1}{2f}\cdot\frac{1}{3h}$, and the move in which
$\faceb$ is the chosen face (if it exists) is made with probability
$\frac{1}{2f}\cdot\frac{1}{3(h+1)}$. Since these moves increase the
distance by $h$ and $h+1$, respectively, we have 
$$\mathbb{E}[\dist_{\facec}] \leq \frac{h}{6fh} + \frac{h+1}{6f(h+1)}
   = \frac{1}{3f}\,.$$

\item In the worst case, we could have a tower starting at $\facec$ in
  both orientations. Let $T_1$ and $T_2$ denote the towers in $\eoi$
  and $\eoii$, respectively, and let $h_1$ and $h_2$ be the lengths of 
  each tower. If $h_1 \leq h_2$ then the coupling reverses the tower
  in both chains with probability $\frac{1}{3h_2}$, and only reverses
  the tower in $\eoi$ with probability
  $\frac{1}{3h_1}-\frac{1}{3h_2}$. The first of these will give two 
  orientations which are distance $h_1+h_2+1$ apart, and the second
  gives two orientations which are $h_1$ apart. Hence, 
  $$\mathbb{E}[\dist_{\facec}] = \frac{1}{2f}\left[
  (h_1+h_2)\frac{1}{3h_2} + h_1(\frac{1}{3h_1}-\frac{1}{3h_2})\right]
  =  \frac{1}{3f}\,.$$  
  The analysis is identical if $h_2 \leq h_1$.  

\end{list}
Combining these two cases, we see that the expected value of
$\dist_{\facec}$ is no more than $\frac{1}{3f}$. 
\end{pf}
\end{lem}

We are now ready to combine the previous lemmas into a proof that the
Markov chain $\mct$ is rapidly mixing.
 
\begin{pf}[Proof of Theorem~\ref{thm:mixing}] Consider the coupling
defined earlier. With probability $1/f$ one of the two chains reverses
$\facea$ causing the two chains to coalesce. Combining this
fact with the results of Lemmas~\ref{lem:directed}
and \ref{lem:undirected} we find that for all $\eoi$ and $\eoii$
differing on the orientation of a single face 
$$\mathbb{E}[\dist(X_{t+1},Y_{t+1})-\dist(X_t,Y_t) | (X_t,Y_t)=(\eoi,\eoii)]
\leq 3\frac{1}{3f} - \frac{1}{f} = 0\,.$$ 
It follows that $\beta=1$ for the path coupling so we can apply
Theorem~\ref{thm:pc} with $p=1/2f$ and
$D\in O(f^{\frac{3}{2}})$ (Lemma~\ref{lem:dist}) to
yield $\tau_{\mct}(\epsilon) \leq Bin(t(\epsilon),(1+p)^{-1})$ where
$$t(\epsilon) \in O(f^4 \log\epsilon^{-1}).$$
But $(1+p)^{-1} \rightarrow 1$ so, for all practical purposes, we can
conclude that $\tau_{\mct} \in O(f^4 \log\epsilon^{-1})$.
\end{pf}

\section{Comparison of $\tau_{\mc}$ and $\tau_{\mct}$}\label{sec:comp}
In this section we use the comparison method of Diaconis \&
Saloff-Coste \cite{DiaconisS-C:AnnalsAppProb:93} to infer a bound on
the mixing time of $\mc$ from the bound on the mixing time of $\mct$
obtained in the previous section. We will use the formulation of the
Diaconis \& Saloff-Coste result from \cite{RandallT:JMP41}, restated
here for convenience. Note that we are using $E(P)$ to denote the set of
edges corresponding to moves between adjacent states in the Markov
chain with transition matrix $P$. 
\begin{thm}\label{thm:comparison}\emph{(\cite[Proposition
      4]{RandallT:JMP41})} 
Suppose $P$ and $\widetilde{P}$ are the transition matrices of two
reversible Markov chains, $\mc$ and $\widetilde{\mc}$, both with the state
space $\Omega$ and stationary distribution $\pi$, and let
$\pi_{\star}=\min_{x \in \Omega}\pi(x)$. For each pair $(u,v) \in
E(\widetilde{P})$, define a path $\path_{uv}$ which is a sequence of
states $u=u_0,u_1,,\ldots,u_k=v$ with $(u_i,u_{i+1}) \in E(P)$ for all
$i$. For $(x,y) \in E(P)$, let $\Gamma(x,y)=\{(u,v) \in
E(\widetilde{P}): (x,y) \in \path_{uv}\}$. Let 
$$
A = \displaystyle\max_{(x,y) \in E(P)} 
\left\{
\frac{1}{\pi(x)P(x,y)}\sum_{(u,v) \in \Gamma(x,y)}
|\path_{uv}|\pi(u)\widetilde{P}(u,v)
\right\}\,.
$$  
Suppose that the second largest eigenvalue, $\lambda_1$, of $\widetilde{P}$
satisfies $\lambda_1 \geq 1/2$. Then for any $0 < \epsilon < 1$

$$\tau_{\mc}(\epsilon) \in
O(A\tau_{\tilde{\mc}}(\epsilon)\log{1/\pi_{\star}})$$ 
\end{thm}  

To use the comparison method to bound the mixing time of the
face-reversal chain we need to show
that every move of $\mct$ can be simulated by moves of the chain $\mc$. Suppose
$T=\{\facea_1,\facea_2,\ldots,\facea_h\}$ is a tower in $\eoi$ and that
$\eoii$ is the orientation obtained by reversing $T$. Observe that by
the definition of a tower $\facea_h$ is a directed cycle in
$\eoi$. Then we can perform a move on $\eoi$ in $\mc$ to obtain a
new Eulerian orientation $\eoi^\prime$ in which $\facea_h$ has been
reversed. But there is now a tower
$T^\prime=\{\facea_1,\facea_2,\ldots,\facea_{h-1}\}$ in $\eoi^\prime$
which can be reversed to obtain $\eoii$. Repeating this process until
we reach $\eoii$ gives a decomposition of the tower move into moves of 
the chain $\mc$. We begin by bounding the size of any tower move:

\begin{lem}\label{lem:maxtower}
Let $G$ be a solid subgraph of the triangular lattice. Then the
maximum length of a tower in an Eulerian orientation of $G$ is
$O(\sqrt{f})$.   
\begin{pf}
Let $T=(\facea_1,\facea_2,\ldots,\facea_h)$ denote a tower in any
Eulerian orientation of $G$. Assume w.l.o.g that $T$ is a clockwise
tower. Recall that all the internal edges
of a clockwise tower are directed towards the vertices on the left, so
any Eulerian orientation of $G$ must contain a set of $h-1$
edge-disjoint directed paths linking the left-side vertices to the
right-side vertices. Each of these paths must go around the top or the
bottom of the tower. But each path that goes around the bottom
(resp. top) contributes $1$ to the distance from $\facea_1$
(resp. $\facea_h$) to the boundary. Hence, by (\ref{expr:pot1}),
(\ref{expr:pot2}) and (\ref{expr:pot3}),
$\max(\pot_{\max}(\facea_1),\pot_{\max}(\facea_h)) \geq h/2$. But
$\pot_{\max}(\facea) \in O(\sqrt{f})$ for any $\facea$ (see proof of
Lemma~\ref{lem:dist}), whence $h \in O(\sqrt{f})$.       
\end{pf}
\end{lem}
We are now ready to prove our rapid mixing result for $\mc$.
\begin{thm}\label{thm:face_mixing}
Suppose $G$ is a solid subgraph of the triangular lattice. Then the
mixing time of the face-reversal Markov chain $\mc$ on $EO(G)$ satisfies
$$\tau_{\mc}(\epsilon) \in O(f^6\log\epsilon^{-1})\,.$$
\begin{pf} Let $P$ and $\widetilde{P}$ denote the transition matrices of $\mc$
and $\mct$. For each pair of states $(x,y)$ that differ
on the orientation of exactly one face (ie. each $(x,y) \in
\textnormal{ker}(\mc)$), we define $\Gamma(x,y)$ to be the set of all
transitions in $\mct$ containing the transition $t=(x,y)$ as a
sub-move. For each such pair we have 
\begin{align}
A_{x,y}&=\frac{1}{\pi(x)P(x,y)}
  \sum_{(u,v)\in\Gamma(x,y)}|\path_{uv}|\pi(u)
  \widetilde{P}(u,v)\\
  &=  2f \sum_{(u,v)\in\Gamma(x,y)} |\path_{uv}|
  \widetilde{P}(u,v)\,, \label{dropprobs}\\
  &= 1 + \frac{1}{3}(|\Gamma(x,y)|-1)\,,\label{dropsum}
\end{align}
where (\ref{dropprobs}) is due to the fact that all transition
probabilities in $P$ are $\frac{1}{2f}$ and that $\pi$ is uniform, and
(\ref{dropsum}) is due to the fact that $\widetilde{P}(u,v)=\frac{1}{2f}$
if $(u,v)=(x,y)$, and $\widetilde{P}(u,v)=\frac{1}{6f|\path_{uv}|}$ if
$(u,v)$ is the reversal of a tower.    

Let $\faceb$ be the face that is reversed in  the transition $t=(x,y)$.
We need to consider the different cases
in which $t$ can feature as part of the decomposition of a tower move
$(u,v) \in E(\widetilde{P})$. Observe 
that there are three different directions in which a tower can pass
through $\faceb$ and contain $t$ as a sub-move (one for each pair of
edges of $\faceb)$. Let $\facea$ and $\facea^\prime$ denote the top
and bottom of the maximal tower (in any Eulerian orientation) which
passes through $\faceb$ in the i-th direction and whose encoding
contains $t$. We use $h_i$ to denote the length of this tower. Any other
tower which passes through $\faceb$ in this direction and whose
encoding uses $t$ must be subset of the faces of the maximal  
tower. Moreover the top (resp. bottom) must lie between $\facea$
(resp. $\facea^\prime$) and $\faceb$. Hence, the number of tower moves
that use a particular transition is $O(h_0^2)+O(h_1^2)+O(h_2^2)$. But
$h_i \in O(\sqrt{f})$ whence $\Gamma(x,y)$ is in $O(f)$ for all $(x,y)
\in \textnormal{ker}(\mc)$. 
 
Thus, we have $A_{x,y} \in O(f)$ for all $(x,y) \in
\textnormal{ker}(\mc)$. The number of edges in $G$ is no more than
$3f$, so $2^{3f}$ provides an upper bound on the number of
orientations of $G$, and so
also on $1/\pi_{\star}$. Combining all this with
Theorems~\ref{thm:mixing} and \ref{thm:comparison} we get   
$$\tau_{\mc}(\epsilon) \in O(f^6\log\epsilon^{-1})\,.$$ 
\end{pf}
\end{thm}

\section{Subgraphs with holes}\label{sec:torpid}

\begin{figure}[ht]
\begin{center}
\includegraphics[height=10cm,width=10cm]{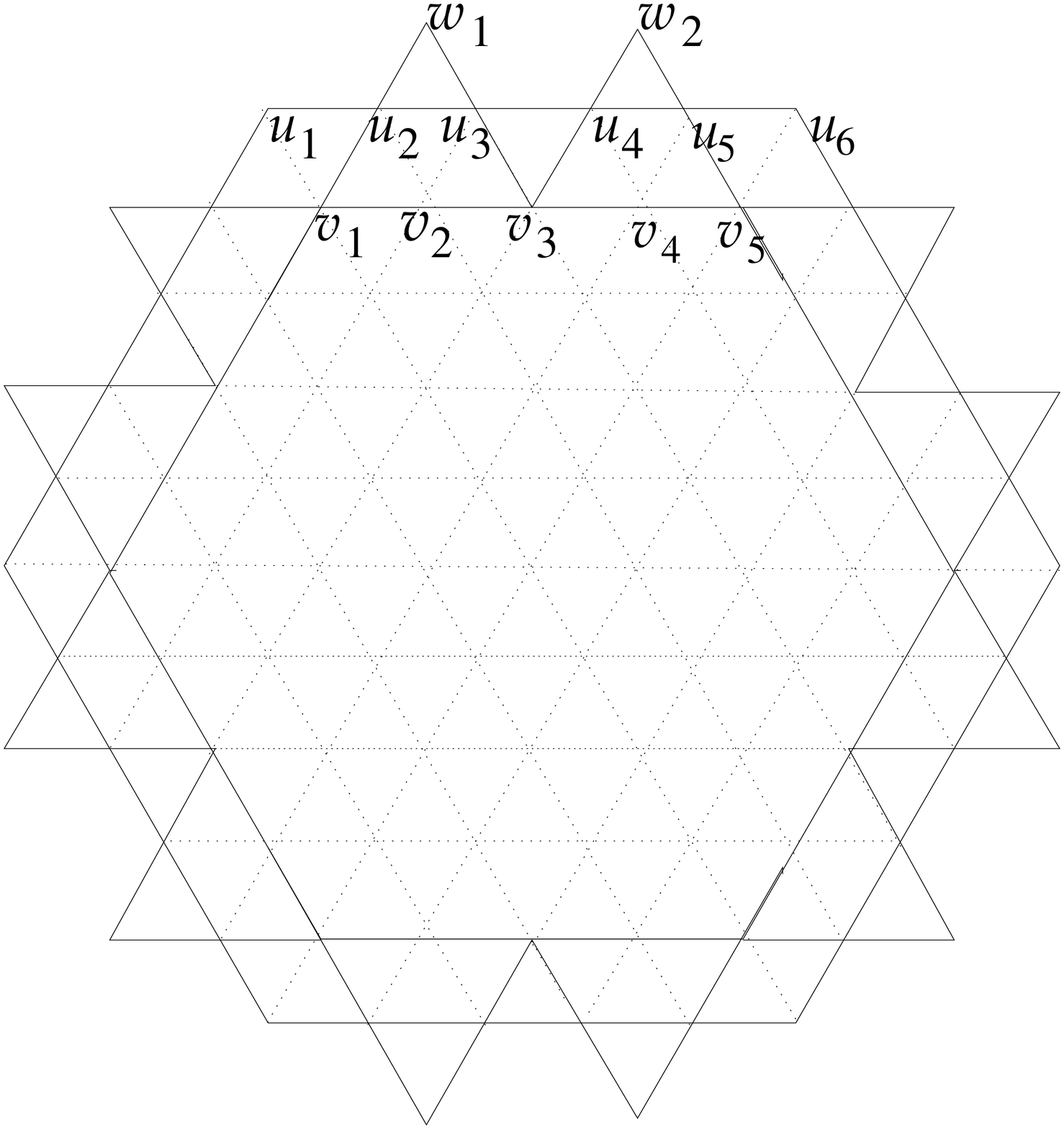}
\end{center}
\caption{The graph $H_2$}\label{fig:h2}
\end{figure}

Given the small collection of positive results regarding the mixing
time of the face-reversal chain $\mc$ (Theorem~\ref{thm:face_mixing}
and \cite{LubyRS:SIAMCOMP01,GoldbergMP:RSA04}) and
given that the reduction of \cite{MihailW:ALGO96}) allows us to sample
from $EO(G)$ in polynomial time for any graph, one  
might hope that $\mc$ might be rapidly mixing on the set of
Eulerian orientations of any planar graph. In fact, this is not true
and in this section we exhibit an infinite family of subgraphs of the
triangular lattice (with holes) for which $\mc$ is torpidly
mixing. Consider the family of graphs $H_N$, of which $H_2$ is shown
in Figure~\ref{fig:h2}. Note that the dotted lines in
Figure~\ref{fig:h2} represent the parts of the triangular lattice
which have been omitted. Formally, $H_N$ is a graph with vertex set   
$$V=\{\vertexv_i:1 \leq i \leq 6(2N+1)\} \cup \{\vertexu_i: 1 \leq i
  \leq 6(2N+2)\} \cup  \{\vertexw_i: 1 \leq i \leq 6N\}\,.$$
The edges of $H_N$ consists of the disjoint union of three large
cycles: 
\begin{align*}
E_1 &= (\vertexv_1,\ldots,\vertexv_{6(2N+1)},\vertexv_1)\\
E_2 &= (\vertexu_1,\ldots,\vertexu_{6(2N+2)},\vertexu_1)\\
E_3 &= (\vertexv_1,\vertexu_2,\vertexw_1,\vertexu_3,\vertexv_3,\ldots,\vertexu_{6(2N+2)},\vertexv_1) 
\end{align*}
eg. see Figure~\ref{fig:h2}. It is the large face in the centre of
each of these graphs that creates the bottleneck in the Markov chain
we will use to show torpid mixing. We label this face $C$ and its
neighbours $\facea_i$ (for $1 \leq i \leq 6N$). The face that  is
  adjacent to both $\facea_i$ 
and $\facea_{i+1}$ is labelled $\faceb_i$, and the face that is only
adjacent to $\facea_i$ is labelled $\facec_i$.

\begin{thm}
The face-reversal chain $\mc$ is torpidly mixing on $EO(H_N)$ for $N
\geq 3$. 
\begin{pf}
From Theorem~\ref{thm:torpid} we know that $\mc$ is torpidly mixing on
a set of Eulerian orientations $\Omega$ if there exists some $S
\subset \Omega$, with $0 < |S| \leq |\Omega|/2$, such that $|\partial
S|/|S|$ is exponentially small in $f$.   

Let $S$ be the set of all Eulerian orientations $\eo$ on $H_N$ satisfying 
$\pot_{\eo}(C) \leq 1$. We can define a bijection between $S$ and
$\Omega \backslash S$ by mapping $\pot_{\eo}$ to
$\pot_{\max}-\pot_{\eo}$ for each $\eo \in S$, so $|S|=|\Omega|/2$.  

An Eulerian orientation $\eo$ is an element of $\partial S$ if and
only if $C$ is a counter-clockwise directed cycle in $\eo$ and
$\pot_{\eo}(C)=1$. For this to occur we must have
$\pot_{\eo}(\facea_i)=1$ for each $\facea_i$. Hence, the number of
Eulerian orientations satisfying this condition is exactly $2^{2k}$
since each of the $\facea_i$ and $\facec_i$ can take potential value
$0$ or $1$, where $k=2N$.  

$C$ is the only directed cycle in $\eo_{\min}$ so $|S|=|S^\prime|+1$, where
$S^\prime=\{\eo \in \Omega: \pot_{\eo}(C)=1\}$. We can partition
$S^\prime$ as $\bigcup_{I \subset [k]}S_I$, where 
$$S_I = \{\eo \in \Omega: \pot_{\eo}(C)=1\,\wedge\,
\pot_{\eo}(\facea_i)=1\,\Leftrightarrow i \in I\}\,.$$
We can find the size of each of the $S_I$ by counting the number of
potential functions which correspond to members of $S_I$. If $\eo \in S_I$
then there are two possible values for $\pot_{\eo}(\faceb_i)$ for each $i$
with $\pot_{\eo}(\facea_i)=1$ and $\pot_{\eo}(\facea_{i+1})=1$, and two
possible values for $\pot_{\eo}(\facec_i)$ for each $i$ with
$\pot_{\eo}(\facea_i)=1$. All of the other $\faceb_i$ and $\facec_i$ must have
potential value $0$. Hence, 
$$|S_I| = 2^{|I|+c(I)}\,,$$
where $c(I)$ counts the number of circular successions in $I$. The
number of $j$-subsets of $\{1,\ldots,k\}$ containing $m$ circular
successions is given by the following expression:
$$c(k,j,m) = \begin{cases}
             0 & \mbox{if }j=0\,, j>k\,, \mbox{ or } m < 2j-k\\
             \frac{k}{j}\binom{j}{m}\binom{k-j-1}{j-m-1} & \mbox{otherwise} 
             \end{cases}\,.
$$ 
Then, 
\begin{align}
|S| &= 1 + \sum_{j=0}^{k}2^j\sum_{m=0}^{j}2^mc(k,j,m)\\
    &=\sum_{j=1}^{k-1}\sum_{m=\max(0,2j-k)}^{j-1}
    \frac{k}{j}\binom{j}{m}\binom{k-j-1}{j-m-1}2^{j+m} + 1 + 2^{2k}\\
    &> \sum_{j=1}^{k-1}\binom{j}{2j-k}2^{3j-k}\quad\\
    &> \binom{\lfloor\frac{16}{17}k\rfloor}{\lceil\frac{1}{17}k\rceil}
       2^{\frac{31}{17}k-3}\quad \mbox{ if } k \geq 17 \\ 
    &\geq 2^{(2+\frac{1}{17})k-3} 
\end{align}      
The last line of this follows from the fact that
$\binom{\lfloor\frac{16}{17}k\rfloor}{\lceil\frac{1}{17}k\rceil} \geq
2^{\frac{4k}{17}}$ when $k \geq 17$.
Hence, 
$$|\partial S|/|S| < 8\cdot2^{-\frac{1}{17}k} \in O(2^{-\frac{1}{51}f})\,.$$ 
 
\end{pf}
\end{thm}

\section{Complexity of \#PlanarEO}\label{sec:planarhard}
In this section we demonstrate a polynomial-time reduction from \#EO
to \#PlanarEO:

\begin{tabular}{ll}
{\bf Name.} & \#EO\\
{\bf Input.} & A graph $G$\\
{\bf Output.} & The number of Eulerian orientations of $G$
\end{tabular}
\\\\
\noindent
\begin{tabular}{ll}
{\bf Name.} & \#PlanarEO\\
{\bf Input.} & A planar graph $G$\\
{\bf Output.} & The number of Eulerian orientations of $G$
\end{tabular}\\

This suffices to show that \#PlanarEO is \#P-complete since the
\#P-completeness of \#EO is already known \cite{MihailW:ALGO96}. Our
reduction uses a recursive gadget and can be seen as an application of
the so-called Fibonacci method of Vadhan \cite{Vadhan:SIAMCOMP01}.

\begin{figure}[ht]
\begin{center}
\begin{tabular}{cc}
\includegraphics[width=4cm]{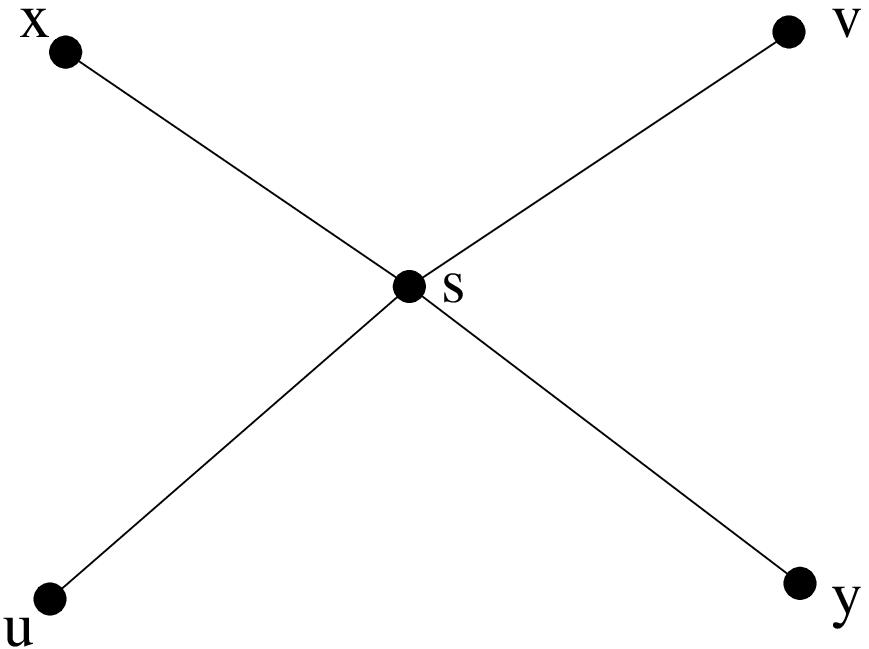} & 
\includegraphics[width=4cm]{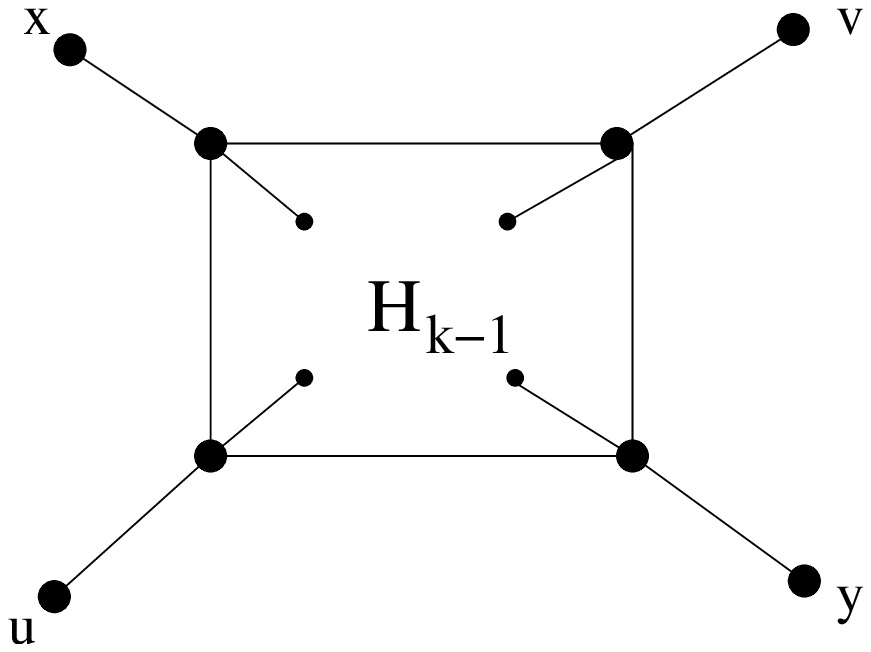} 
\\
(a) & (b) 
\end{tabular}
\end{center}
\caption{The crossover box}
\label{fig:crossover}
\end{figure} 

\begin{thm}\label{thm:peohard}
There exists a polynomial time reduction from \#EO to \#PlanarEO.
\begin{pf}
Let $G$ be any non-planar graph for which we have an embedding in the plane
with $l$ crossings. We start by creating a planar graph $G^\prime$ by
replacing each crossing $\{x,y\}$ and $\{u,v\}$ by a vertex $s$ joined
to each of $u,v,x,y$ as in Figure \ref{fig:crossover}(a). Using our
oracle for \#PlanarEO we can count the number of Eulerian orientations
of $G^\prime$ in polynomial time. Not all of these correspond
to Eulerian orientations of $G$, eg. $\{(u,s),(v,s)\}$ may be present
in an Eulerian orientation of $G^\prime$. We call the configurations
of arcs at a crossover that correspond to Eulerian orientations of $G$
\emph{valid} configurations.

For each $k$ we define $H_k$ recursively as in
Figure~\ref{fig:crossover}(b) with $H_0$ given by
Figure~\ref{fig:crossover}(a). Let $G_k$ be the graph obtained by
replacing each crossing in $G$ with $H_k$, so $G_0=G^\prime$.  

Now let $x_k$ (resp. $y_k$) denote the number of possible
configurations of the edges of $H_k$ satisfying the Eulerian condition
which correspond to \emph{valid} (resp. \emph{invalid}) orientations
of $G^\prime$. These values satisfy 
\begin{eqnarray}
x_k &= 4x_{k-1} + 2y_{k-1}\\
y_k &= 4x_{k-1} + 3y_{k-1}
\end{eqnarray}
with $x_0=y_0=1$. Now let $N_{i}$ denote the number of Eulerian
orientations of $G^\prime$ which have exactly $i$ valid crossover
boxes, so $N_l=\#EO(G)$. Each Eulerian orientation of $G^\prime$
counted by $N_i$ corresponds to exactly $x_k^iy_k^{l-i}$ Eulerian
orientations of $G_k$, so we can write
$$\#PlanarEO(G_k) = \sum_{i=0}^{l}N_ix_k^iy_k^{l-i}\,.$$
Then calculating $\#PlanarEO(G_k)/y_k^l$ corresponds to evaluating
the polynomial $p(z) = \sum_{i=0}^{l}N_iz^i$ at the point
$x_k/y_k$. Since $x_k/y_k$ is a non-repeating sequence (see
\cite[Lemma 6.2]{Vadhan:SIAMCOMP01}) it suffices to evaluate
$\#PlanarEO(G_k)/y_k^l$ for $k=0\ldots l+1$ to obtain enough information
to recover the values of $N_i$ by polynomial interpolation. 

Letting $n$ and $m$ denote the number of vertices and edges in $G$ we
have $|V(G_k)|=n+(4k+1)l$ and $|E(G_k)|=m+(8k+2)l$. Since $l$ is certainly
bounded by a polynomial in $n$ it follows that the whole reduction can
be performed in time polynomial in $n$ and $m$.
\end{pf}
\end{thm}

\section{Conclusions}
We have shown that the mixing time of the face-reversal chain is
$O(f^6)$ on the set of Eulerian orientations of any solid subgraph of
the triangular lattice. This compares favourably with
$O(f^7\log^4(f))$, the best known bound on the sampling algorithm for
general graphs \cite{MihailW:ALGO96,BezakovaSVV:SODA06}. On the other
hand we have demonstrated that there exist subgraphs of the triangular
lattice for which the face-reversal chain takes an exponential amount
of time to converge. Given that the presence of a single very large
face is the obstacle to rapid mixing in these graphs one might ask
whether there exists some function $g(f)$ such that the face-reversal
chain mixes rapidly whenever the number of edges in any face is
bounded by $g(f)$. It seems that this question is beyond current proof
techniques although a proof based on analysing the conductance of cuts
similar to that used in \S\ref{sec:torpid} seems plausible. 
 
We also showed that the problem of finding an exact value for
$|EO(G)|$ is \#P-complete when $G$ is a planar graph. This does not
preclude a polynomial-time algorithm for evaluating $|EO(G)|$ for more
restricted classes of graphs, eg. solid subgraphs of the triangular
lattice. However, if such an algorithm cannot be found then our rapid
mixing result could be used to construct a \emph{fully polynomial
randomised approximation scheme}, that is, an algorithm which can
approximate the value of $|EO(G)|$ to arbitrary precision $\epsilon$
and runs in time polynomial in $f$ and $\epsilon^{-1}$. We
sketch such an algorithm below:  

Let $G$ be a solid subgraph of the triangular lattice and let $\facea$
be a face on the boundary  of $G$. We construct $G_1$ by deleting the
edges of $\facea$ from $G$ and removing any isolated vertices. It is
straightforward to check that $|EO(G_1)|/|EO(G)| \geq 1/4$. Hence, we
can approximate this value to arbitrary precision with relatively few
samples from $EO(G)$. But $G_1$ is also a solid subgraph of  
the triangular lattice so we can repeat this procedure until 
we reach a final $G_k$ for which we can easily find $|EO(G_k)|$,
eg. $K_3$. Since the number of faces decreases by at least $2$ at each 
step of this process it follows that $k \leq f/2$. Hence, we can
estimate $|EO(G)|$ to arbitrary precision $\epsilon$ in time
polynomial in $f$ and in $\epsilon^{-1}$. A detailed description of
this general method to convert efficient samplers into approximate
counting algorithms can be found in \cite{DyerG:LMS267}.   
\section*{Acknowledgements}
This work benefited from useful conversations with Mary Cryan and Mark
Jerrum. I am especially grateful to Mary Cryan for her many comments
on an early version of this paper. I would also like to thank Magnus
Bordewich for providing me with a preprint of
\cite{BordewichD:Equality}, and Florian Zickfeld for comments on a
preliminary version of Theorem~\ref{thm:peohard}. 

\bibliographystyle{elsart-num-sort}
\bibliography{eulerian}

\end{document}